%% file: async-ckp.tex
\pdfoutput=1
\documentclass{article}

 \usepackage[nonatbib, preprint]{nips_2018}
\usepackage[utf8]{inputenc} 
\usepackage[T1]{fontenc}    
\usepackage{hyperref}       
\usepackage{xcolor}         
\usepackage{graphicx}
\usepackage{url}            
\usepackage{booktabs}       
\usepackage{amsfonts}       
\usepackage{nicefrac}       
\usepackage{microtype}      
\usepackage{amsmath}
\usepackage[numbers]{natbib}
\title{Backpropagation for long sequences: beyond memory constraints
  with constant overheads}

\author{Navjot Kukreja \\
  Department of Earth Science and Engineering\\
  Imperial College London\\
  \texttt{nkukreja@imperial.ac.uk} \\
  \And
  Jan H\"uckelheim \\
  Department of Earth Science and Engineering\\
  Imperial College London\\
  \texttt{j.hueckelheim@imperial.ac.uk} \\
  \And
  Gerard J. Gorman \\
  Department of Earth Science and Engineering\\
  Imperial College London\\
  \texttt{g.gorman@imperial.ac.uk} \\
}

\graphicspath{ {./figures/} }

\begin{document}

\maketitle

\begin{abstract}
Naive backpropagation through time has a memory footprint that grows
linearly in the sequence length, due to the need to store each state
of the forward propagation. This is a problem for large
networks. Strategies have been developed to trade memory for added
computations, which results in a sublinear growth of memory footprint
or computation overhead. In this work, we present a library that uses
asynchronous storing and prefetching to move data to and from slow and
cheap storage.  The library only stores and prefetches states as
frequently as possible without delaying the computation, and uses the
optimal Revolve backpropagation strategy for the computations in
between. The memory footprint of the backpropagation can thus be
reduced to any size (e.g. to fit into DRAM), while the computational
overhead is constant in the sequence length, and only depends on the
ratio between compute and transfer times on a given hardware. We show
in experiments that by exploiting asyncronous data transfer, our
strategy is always at least as fast, and usually faster than the
previously studied ``optimal" strategies.
\end{abstract}

\section{Introduction}
The current trend is towards training ever deeper networks as deeper
networks have a larger capacity to learn. Since backpropagation
requires the complete state of the forward propagation in reverse
order, training a neural network with backpropagation requires memory
that is proportional to the size of the network. Many state-of-the-art
models already run out of memory on current hardware and this trend is
only expected to get worse. \cite{rhu2016vdnn}

One of the most common ways of managing memory consumption of neural
network training is by controlling the batch size
\cite{rhu2016vdnn}. However, since the batch size is also used to
sample from the training data, the choice of batch size can affect the
convergence rate and cannot be used to tune the model's memory
consumption without side-effects.

Another common mitigation strategy is to split the training over
multiple computational nodes \cite{krizhevsky2014one}.  However, this
incurs significant message passing overheads and costs for hardware
with low-latency interconnects.  This strategy can also be wasteful if
the peak memory consumption is only slightly larger than that of a
single compute node.

A third strategy that is recently getting increased attention is checkpointing,
and is briefly reviewed in the following section.

\subsection{Checkpointing for neural networks}
\label{sec:intro}

\begin{figure}
\begin{center}
\def\svgwidth{0.5\textwidth}
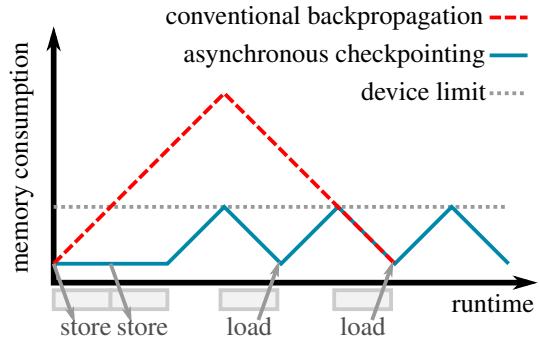
\end{center}
\caption{Memory requirement of a neural network during training. In conventional
backpropagation, all states need to be stored, leading to a peak memory
footprint at the end of the forward computation. During the backward pass, the
stored states are used and their memory subsequently freed in the reverse order.
Training can not be performed on hardware with too small memory. In contrast,
checkpointing strategies store some intermediate states and resume recomputation
from there when required. With asynchronous multistage checkpointing, the data
is further offloaded to a larger, slower storage system (e.g. solid state drive)
in the background while the computation is running, and prefetched before it is
needed.}
\label{fig:title}
\end{figure}

The idea behind checkpointing is not to store the entire state of the
network through the forward propagation. Instead, the state of forward
propagation is stored only at certain layers, and the number of layers
that are kept at any given time can be limited to fit into the
available memory.  During the backpropagation phase, states that have
not been stored can be recomputed as needed from the nearest available
state.  This allows a tradeoff between memory and computation. With
this, problems can be made to fit on systems with limited memory in
exchange for an increased computation time.

The pressure on the memory system during a backpropagation
execution can be quantified using a \emph{memory ratio}, i.e. the
ratio between the memory available on a computer system and the
expected peak memory consumption of a particular instance of
backpropagation. We are only interested here in scenarios where the
memory ratio is less than 1.

The amount of recomputation required in a checkpointing strategy is
quantified using a \emph{recompute factor} where a factor of 1 implies
no recomputation. The factor grows as the memory ratio is reduced. The
choice of layers at which to store checkpoints during the forward
propagation directly affects the recompute factor and is called the
\emph{checkpointing schedule}.

Checkpointing is widely used for similar purposes in adjoint based
optimisation problems, and a number of schedules have been developed
that are optimal under certain assumptions. If the number of layers is
known a priori, states have a uniform size, each layer takes the same
time to compute, and the memory is fast and the time to store a
checkpoint is thus negligible, then the Revolve algorithm
\cite{griewank2000algorithm} gives the optimal schedule that minimises
recomputation given a fixed amount of memory. Another schedule has
been found to be optimal if the number of layers is not known
initially \cite{wang2009minimal}. The development of these algorithms
was motivated by adjoint solvers, where these assumptions are usually
valid.

In contrast, the state size and computation cost of layers in neural
networks is often non-uniform (e.g. different before and after a
pooling layer). New checkpointing schedules have been developed
specifically for machine learning
applications~\cite{chen2016training}, including a dynamic program that
can be used to compute optimal schedules for networks of uniform and
non-uniform checkpoint sizes~\cite{gruslys2016memory}.

\subsection{Multistage checkpointing}

When additional levels of memory are available, it is possible to
leverage these additional levels to reduce the recompute factor
\cite{stumm2009multistage}.  In the context of modern computer
systems, the two levels of memory could be the accelerator memory and
the host memory. Even on systems where only one level of memory is
usually used, the second level memory could be a high-bandwidth disk,
e.g. an SSD. In the foreseeable future, other types of memory are
expected to become available, such as storage-class
memory~\cite{JSFI164}.

For systems with two levels of memory, \cite{aupy2016optimal}
describes the optimal schedule that reduces the total time to solution
for adjoint solvers or backpropagation, assuming that the first level
memory is fast but has limited-capacity, while the second level is
slow but has infinite capacity. The key idea is to increase the number
of stored checkpoints, by storing the least frequently used
checkpoints on the slow, large storage. The schedule assumes blocking
data transfer, that is, the computation waits while data is
transferred from the fast to the slow storage level.

Since transfers between first-level memory and second-level memory
take a non-trivial amount of time, they can be carried out in
parallel. This motivated a recent paper~\cite{schanen2016asynchronous}
describing the use of asychronous multistage checkpointing for a PDE
solver. In that work, the solver itself uses all available RAM on a
system, and the checkpoints are thus stored directly to a hard
drive. Since the overall stored data is much larger than available
hard drives, another system is transferring the data over the network
to a tape archive while the computation is running.

A similar concept was also previously applied to neural networks
\cite{rhu2016vdnn}. However, in this case every layer was transferred
to the second-level memory, which slows down the forward
propagation. A variation of this strategy, where a subset of states is
transferred to the host memory and transferred back when required was
also implemented for Tensorflow, but without any recomputation of
forward propagated states. \cite{mengtraining}

\subsection{Contributions}

While the work in this paper is conceptually similar to that presented
in~\cite{schanen2016asynchronous}, to the best of our knowledge,
multistage checkpointing with recomputation of forward states has not
been applied in the context of neural networks before. It has also not
previously been investigated for systems other than the aforementioned
hard drive/tape system. This is despite the fact that non-blocking
asynchronous data transfer is possible on a variety of commonly used
systems, such as GPU DRAM / CPU RAM, or from host RAM to another
devide using direct memory access (DMA).  We therefore investigate
asynchronous multistage checkpointing for neural networks on a system
that consists an Intel XeonPhi Knight's Landing (KNL), where the main
computation and Level 1 memory is in fast MCDRAM, and the Level 2
storage is in the system's DRAM.  Figure~\ref{fig:title} gives a
high-level illustration of this idea.

After presenting the scheme in Section~\ref{sec:method}, we present a
performance model for asynchronous checkpointing that works across a
variety of hardware configurations in Section~\ref{sec:model}. We also
developed a prototype implementation for asynchronous multistage
checkpointing in Python, shown in Section~\ref{sec:implementation}. In
Section~\ref{sec:experiments}, we demonstrate the use of this scheme
on two different modern hardware platforms using an LSTM based network
as a test case. We conclude in Section~\ref{sec:conclusions}.

\section{Asynchronous multistage checkpointing}
\label{sec:method}

In this section we outline the asynchronous multistage checkpointing strategy.
We assume that there are two storage stages: Level 1, a fast but small memory,
and Level 2, a large but slow storage. Examples for Level 1 memory include GPU
DRAM or Xeon Phi MCDRAM, while a example for Level 2 storage is a solid
state drive (SSD). Note that these roles depend on the overall
configuration of the system. For example, RAM could either be a Level 2 storage
in a system that is using DRAM as Level 1, or it could be Level 1 memory in a
system that is using SSD or a hard drive for Level 2. What matters is not the
absolute speed or size of the storage, but rather the relative speed and size
compared to other storage in the same system.

In the asynchronous multistage checkpointing strategy, the computation itself
completely resides in Level 1 memory. During the
forward pass, copies of the state are transferred to the Level 2 storage  at
regular intervals, i.e. after every $I$ layers, where $I$ is the \emph{checkpointing
interval}. The transfer to storage happens asychnronously
so as to not slow down the computation of the forward propagation. All
forward activations are then cleared from Level 1 memory.

The backward pass will require the stored data in reverse order, at well-known
points in time during the computation. For this reason, checkpoints that are
required from Level 2 storage can be asynchronously transferred to Level 1
before they are needed. Since every $I$-th state was stored, the intermediate
states need to be recomputed from the restored state.
Assuming there is enough Level 1 memory available to store the entire
forward propagation state for $I$ layers, backpropagation can then
proceed normally for these $I$ layers. If there is not enough
memory available, Revolve can be applied to find an optimal schedule for
backpropagation through $I$ layers within the limits of the Level 1 memory.

\begin{figure}
\begin{center}
\def\svgwidth{0.8\textwidth}
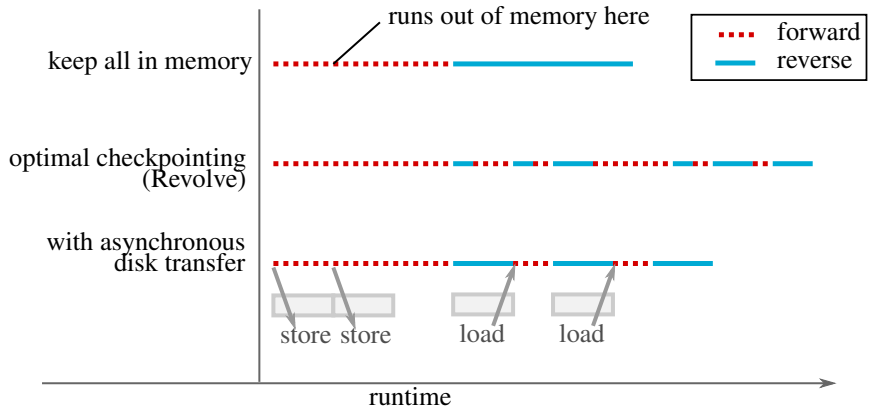
\end{center}
\caption{
Timeline of events for conventional backpropagation, Revolve checkpointing, and
asynchronous multistage checkpointing. The conventional backpropagation would
have the shortest runtime, but exceeds the available memory. Both other
strategies respect the memory limits, but result in different time overheads.
Revolve alternates between forward and reverse computations in a rather complex
fashion to minimise the overhead if only one level of memory is available. The
asynchronous strategy stores data to Level 2 storage in regular intervals, and
restores the data before it is needed in backpropagation.
}
\label{fig:timeline}
\end{figure}

Compared to conventional backpropagation where every state is stored, this
obviously has the advantage that it can fit into limited amounts of memory.
Perhaps less obviously, this strategy is guaranteed to be faster than the
``optimal'' Revolve checkpointing strategy. This is because Revolve (or any of
the other published single-stage checkpointing strategies) trades memory for
additional computations, resulting in a time overhead that increases with the
number of layers. Through the use of Level 2 storage, Revolve is only used for
the $n$ states between two subsequent stores, resulting in a time overhead that
is constant in the number of layers. This is illustrated in
Figure~\ref{fig:timeline} and explained in more detail in
Section~\ref{sec:model}.

\section{Performance Model}
\label{sec:model}

We analyse in this section the expected performance of asynchronous
multistage checkpointing and compare it with Revolve checkpointing. Following
that, we demonstrate the performance in an experiment in
Section~\ref{sec:experiments}.

On a given target computer, let the time taken to compute one layer's
activations be given as $T_A$ and the time taken to propagate sensitivities
backwards through that layer as $T_B$. For a network with $n$ layers, the total
time $T_\infty$ for a combined forward/backward pass as used in training, assuming that
there is no memory limit, is then obviously
$$T_\infty = n\cdot T_A + n\cdot T_B.$$

If Revolve checkpointing is used, some states need to be recomputed,
leading to additional computations of activations. This is expressed
in the \emph{recompute factor}, which depends on the total number of
layers $n$, as well as the number of checkpoints that simultaneously
fit into memory, $s$. We refer to this as
$$R(n,s)$$. The recompute factor is defined in~\cite{revolve}, and can
be computed by the reference implementation of Revolve or by using the
pyrevolve Python package that can be downloaded from
\texttt{https://github.com/opesci/pyrevolve/}. We note that the
recompute factor increases if the number of layers $n$ is increased,
and also increases if the storage space $s$ is decreased. This is true
for all known single-stage checkpointing schemes, and the precise
nature of the increase (sub-linear for most schemes, logarithmic for
Revolve) determines the optimality of a schedule.  The total time
$T_\textit{revolve}$ for the combined forward/backward pass is then
$$T_\textit{revolve} = n\cdot R(n,s) \cdot T_A + n\cdot T_B.$$

For asynchronous multistage checkpointing, we are also interested in the time
that it takes to transfer a state from Level 1 memory to Level 2 storage. We
refer to this time as $T_T$. If $T_T \leq T_A$, then we could asynchronously
stream all data to storage while the computation is running without ever waiting
for disk access. If $T_T > T_A$, then we can only store a subset of all states. We choose
to store states in regular intervals of length $I$, given by
$$I=\left\lceil{\frac{T_T}{T_A}}\right \rceil.$$
In general, there are then $n/I$ such intervals. Storing and prefetching happens
asynchronously, meaning that these operations do not affect performance in
this model (albeit they have a slight effect on performance in practice, see
Section~\ref{sec:experiments}. Within each interval, we can
use Revolve with a recompute factor of $R(I,s)$. Overall, we thus have a runtime
\begin{align*}
T_\textit{async} &= \frac{n}{I}\cdot\left(I\cdot R(I,s)\cdot T_A + I\cdot T_B\right) \\
                 &= n\cdot R(I,s) \cdot T_A + n\cdot T_B.
\end{align*}

\begin{figure}
\begin{center}
\def\svgwidth{0.6\textwidth}
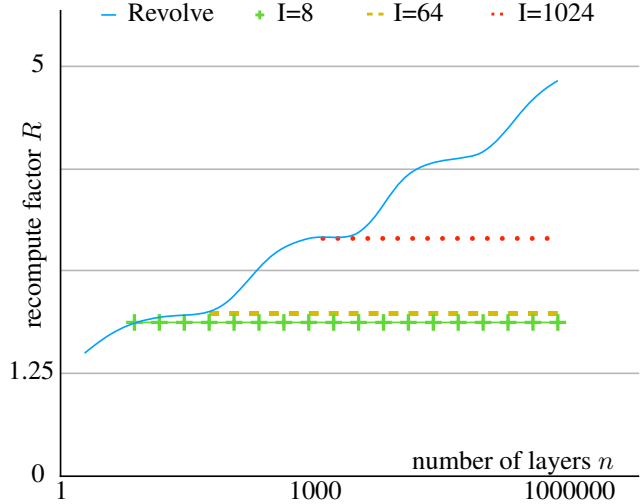
\end{center}
\caption{
Recompute factors, assuming that $s=100$ (that is, $100$ states fit into memory), for classic
Revolve, and asynchronous multistage checkpointing with interval sizes
$I=8,64,1024$.
}
\label{fig:perf}
\end{figure}

Due to the fact that $R(I,s) \leq R(n,s)$ if the interval is at most $n$, the
asynchronous strategy is at least as fast as the classic Revolve strategy. In
particular, the recompute factor in $T_\textit{async}$ depends only on $I$, not
on the total sequence length $n$. Figure~\ref{fig:perf} shows this for a
small number of interval lengths and assuming that $100$ states fit into memory.

Note that in the case where there are very few layers, there might not be time
to save a single checkpoint to second level memory before the entire
forward pass is over. In this case this strategy would fall back to
classic Revolve.

\section{Implementation}
\label{sec:implementation}
The Revolve algorithm was accompanied by a similarly named utility
that could be used to compute optimal schedules for a particular
checkpointing scenario. pyrevolve \cite{kukreja2018high} is a python
package that uses schedules automatically derived from this utility to
provide checkpointing as a feature in python applications with minimal
changes to the code. pyrevolve expects function references to a
\emph{Forward Operator}, and a \emph{Backward Operator}, along with a
\emph{Checkpoint} object that describes the state variables from the
\emph{Forward Operator} that the \emph{Backward Operator}
requires. Provided these three objects, pyrevolve can drive the entire
forward and backward passes, automatically calling the forward or
backward operator as required by the optimal schedule. The
implementation of the asynchronous multistage checkpointing strategy
is offered as an additional mode in pyrevolve
\footnote{https://github.com/opesci/pyrevolve}. Due to the way it has
been formulated, pyrevolve, and consequently the implementation for
this strategy, can be used in applications ranging from
PDE-constrained optimisation problems in seismic imaging and CFD to
neural networks.

The implementation uses the python threading API to create background
threads where the asynchronous reads and writes happen. Python threads
are known to suffer from issues relating to the Global Interpreter
Lock (GIL). However, python releases the GIL when doing IO-bound
operations \cite{pythonGIL}. Hence, this implementation is expected to
be asynchronous despite, if not even due to, the python GIL.

As of now, we implemented this strategy with two hardware
architectures in mind - compute on CPU, DRAM for first level memory
and SSD for second level memory - here we shall call this the CPU
platform. The second architecture is - compute on an accelerator such
as the Intel\textsuperscript{\textregistered} Xeon
Phi\textsuperscript{\texttrademark} 7210 (KNL), with the accelerator
memory, the MCDRAM in the case of the KNL, acting as the first-level
memory and the host memory, or the DRAM in the case of KNL, acting as
the second-level memory. In principle, what we describe here for the
KNL platform applies equally to a GPU architecture where the GDDR
memory acts as the first level and the host memory acts as the second
level.

On the CPU platform, the background threads use the SSD by writing and
reading the data to files using the python filesystem API. On the KNL
platform, a ramdisk mounted to host memory is used as a second level
memory, though this could be improved in future implementations. 

\section{Experimental Results}
\label{sec:experiments}
The test case on which to measure the performance of this strategy and
implementation was adapted from an open source implementation of
simple vanilla LSTM
\footnote{https://github.com/kevin-bruhwiler/Simple-Vanilla-LSTM}. An
LSTM was chosen because a simple LSTM has uniformly sized checkpoints
as we go through the network. Using one of the popular frameworks like
Tensorflow or pyTorch we could have implemented an LSTM in very few
lines but the multiple layers of abstraction involved would hide some
very important details that were relevant for this study. For example,
the framework might be calling precompiled primitives for performant
calculations, and choosing which implementation of a function to call
based on runtime parameters. This caused spikes at certain network
depths that are not relevant to the study at hand. Another issue was
about the transparency of memory management, since we would like to
choose exactly which objects to keep in memory. However, because the
purpose of this experiment is to demonstrate the principle of
asynchronous multistage checkpointing, we believe that this
implementation written with \emph{numpy} as the only external library
is sufficiently representative of a full-fledged LSTM training inside
any of the popular NN frameworks. 

\label{sec:experiments}
\begin{figure}
\begin{center}
\includegraphics[width=0.6\textwidth]{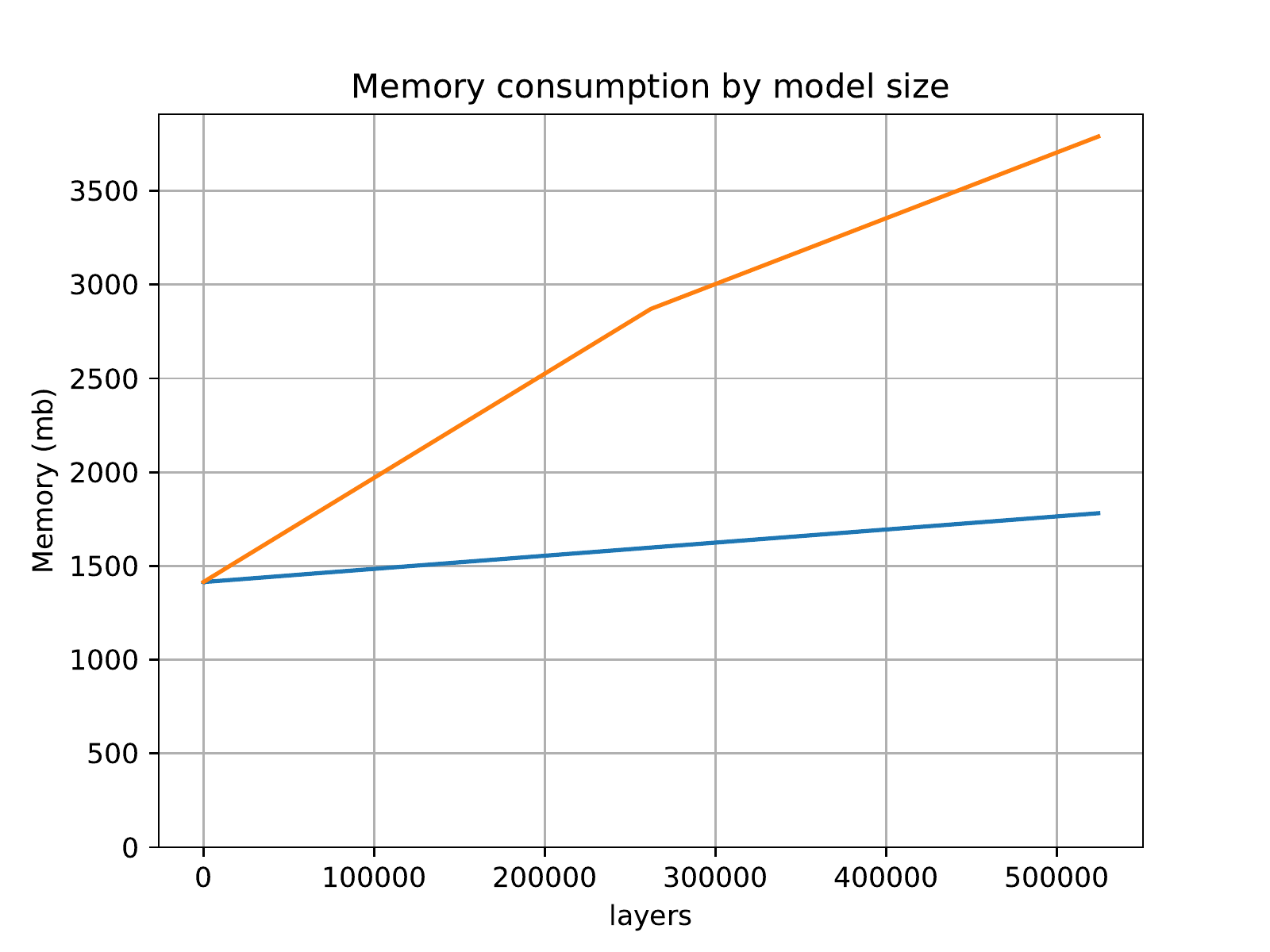}
\end{center}
\caption{Comparison of memory consumption on CPU}
\label{fig:memory}
\end{figure}

The test case \footnote{Code provided as supplementary material} sets
up a basic LSTM for text generation, including a manual implementation
of \emph{RMSProp}. Additional tweaks like learning rate decay would
probably help the convergence of this code in a real-life
scenario. However here we are not concerned about a complete training
cycle, our interest is limited to a single forward-backward iteration
and its performance characteristics as the number of LSTM recurrences
is changed.

\label{sec:experiments}
\begin{figure}
\begin{center}
\includegraphics[width=0.6\textwidth]{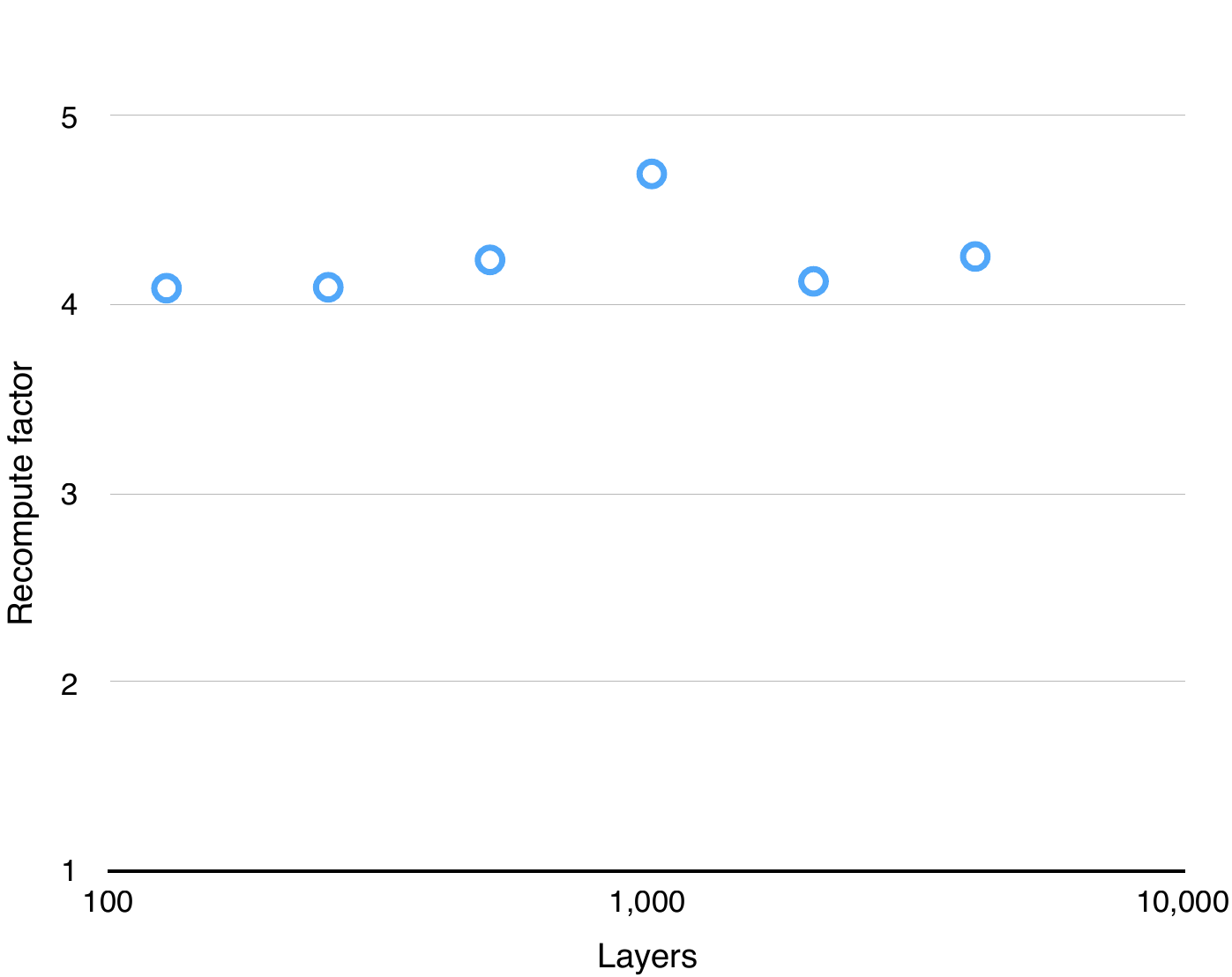}
\end{center}
\caption{Comparison of recompute factors on KNL}
\label{fig:recompute}
\end{figure}

Figure \ref{fig:memory} shows the peak memory footprint for a
single forward-backward pass for a network of given depth, and figure
\ref{fig:recompute} shows how the recompute factor varies with 
network depth. The times
were measured for 5 individual runs and the minimum
reported. The memory reported was measured using the \emph{maximum
  resident set size} reported by the \emph{time} utility on the bash
command line. The python interpreter was exited after each iteration
to ensure that the memory is released back to the OS.

Although the peak memory footprint is theoretically expected to be constant,
regardless of the number of recurrent layers, we observe in the plots
that the memory does go up slightly although at a rate significantly
lower than standard backpropagation. This is because the
implementation still requires some variables whose size is dependent
on the depth of the network. In the case of this LSTM implementation,
the list of expected outputs is the main such variable that can not be
easily made to be independent of the depth of the network.

\section{Conclusions and future work}
\label{sec:conclusions}

We introduced asynchronous multistage checkpointing for backpropagation in large
RNNs in environments with limited memory. The method allows backpropagation
through sequences with arbitrary length for any given memory size, by
recomputing some data and asynchronously transferring other data to a larger,
slower storage such as host memory, RAM, or even SSDs. The runtime overhead
compared to a pure inference is constant in the sequence length, as was shown in
our experiment. The overhead is also at most as large as that of the optimal
single-stage checkpointing strategy Revolve, as shown in a theoretical
performance model.

The implementation currently only supports networks 
that have layers of the same size throughout, i.e. uniform checkpoint
size. Instead of storing every $I$-th state for some fixed interval $I$, one
could instead easily store the next state whenever the previous data transfer
has completed, thereby supporting non-uniform checkpoint sizes.
Within each
interval, the known algorithm for non-uniform single-stage checkpointing could
be used instead of Revolve.

The implementation currently supports Intel XeonPhi processors.
In future work, we plan to extend our implementation to support more platforms,
such as GPUs. Finally, the current implementation assumes that the states within
each interval fit into memory, and this was true for the experiments conducted
in this work. If required, our package can be modified to use Revolve within
each interval, for example using the pyrevolve package.

\subsubsection*{Acknowledgments}

This work has been funded by the Intel Parallel Computing Centre at
Imperial College London. This paper benefitted greatly from
conversations with Paul Kelly, Nicolas Melot, Lukas
Mosser, Paul Hovland and Michel Schanen. This work was performed using
the Darwin Supercomputer of the University of Cambridge High
Performance Computing Service (http://www.hpc.cam.ac.uk/), provided by
Dell Inc. using Strategic Research Infrastructure Funding from the
Higher Education Funding Council for England and funding from the
Science and Technology Facilities Council.


\bibliography{multi_checkpointing}

\end{document}

%% file: figures/async.pdf_tex
\begingroup%
  \makeatletter%
  \providecommand\color[2][]{%
    \errmessage{(Inkscape) Color is used for the text in Inkscape, but the package 'color.sty' is not loaded}%
    \renewcommand\color[2][]{}%
  }%
  \providecommand\transparent[1]{%
    \errmessage{(Inkscape) Transparency is used (non-zero) for the text in Inkscape, but the package 'transparent.sty' is not loaded}%
    \renewcommand\transparent[1]{}%
  }%
  \providecommand\rotatebox[2]{#2}%
  \ifx\svgwidth\undefined%
    \setlength{\unitlength}{235.18558502bp}%
    \ifx\svgscale\undefined%
      \relax%
    \else%
      \setlength{\unitlength}{\unitlength * \real{\svgscale}}%
    \fi%
  \else%
    \setlength{\unitlength}{\svgwidth}%
  \fi%
  \global\let\svgwidth\undefined%
  \global\let\svgscale\undefined%
  \makeatother%
  \begin{picture}(1,0.62161694)%
    \put(0,0){\includegraphics[width=\unitlength,page=1]{async.pdf}}%
    \put(0.66418005,0.44812568){\color[rgb]{0,0,0}\makebox(0,0)[lb]{\smash{device limit}}}%
    \put(0.29195764,0.59201419){\color[rgb]{0,0,0}\makebox(0,0)[lb]{\smash{conventional backpropagation}}}%
    \put(0.32609996,0.51986006){\color[rgb]{0,0,0}\makebox(0,0)[lb]{\smash{asynchronous checkpointing}}}%
    \put(0.8367207,0.0436433){\color[rgb]{0,0,0}\makebox(0,0)[lb]{\smash{runtime}}}%
    \put(0.02960288,0.11824949){\color[rgb]{0,0,0}\rotatebox{90}{\makebox(0,0)[lb]{\smash{memory consumption}}}}%
    \put(0,0){\includegraphics[width=\unitlength,page=2]{async.pdf}}%
    \put(0.09090158,0.0004967){\color[rgb]{0.30196078,0.30196078,0.30196078}\makebox(0,0)[lb]{\smash{store}}}%
    \put(0,0){\includegraphics[width=\unitlength,page=3]{async.pdf}}%
    \put(0.19913262,0.0004967){\color[rgb]{0.30196078,0.30196078,0.30196078}\makebox(0,0)[lb]{\smash{store}}}%
    \put(0,0){\includegraphics[width=\unitlength,page=4]{async.pdf}}%
    \put(0.4076138,0.00060897){\color[rgb]{0.30196078,0.30196078,0.30196078}\makebox(0,0)[lb]{\smash{load}}}%
    \put(0,0){\includegraphics[width=\unitlength,page=5]{async.pdf}}%
    \put(0.6240759,0.00060897){\color[rgb]{0.30196078,0.30196078,0.30196078}\makebox(0,0)[lb]{\smash{load}}}%
  \end{picture}%
\endgroup%

%% file: figures/asynctimeline.pdf_tex
\begingroup%
  \makeatletter%
  \providecommand\color[2][]{%
    \errmessage{(Inkscape) Color is used for the text in Inkscape, but the package 'color.sty' is not loaded}%
    \renewcommand\color[2][]{}%
  }%
  \providecommand\transparent[1]{%
    \errmessage{(Inkscape) Transparency is used (non-zero) for the text in Inkscape, but the package 'transparent.sty' is not loaded}%
    \renewcommand\transparent[1]{}%
  }%
  \providecommand\rotatebox[2]{#2}%
  \ifx\svgwidth\undefined%
    \setlength{\unitlength}{356.625bp}%
    \ifx\svgscale\undefined%
      \relax%
    \else%
      \setlength{\unitlength}{\unitlength * \real{\svgscale}}%
    \fi%
  \else%
    \setlength{\unitlength}{\svgwidth}%
  \fi%
  \global\let\svgwidth\undefined%
  \global\let\svgscale\undefined%
  \makeatother%
  \begin{picture}(1,0.4717309)%
    \put(0,0){\includegraphics[width=\unitlength,page=1]{asynctimeline.pdf}}%
    \put(0.30148977,0.07447168){\color[rgb]{0.30196078,0.30196078,0.30196078}\makebox(0,0)[lb]{\smash{store}}}%
    \put(0,0){\includegraphics[width=\unitlength,page=2]{asynctimeline.pdf}}%
    \put(0.37276544,0.07447204){\color[rgb]{0.30196078,0.30196078,0.30196078}\makebox(0,0)[lb]{\smash{store}}}%
    \put(0,0){\includegraphics[width=\unitlength,page=3]{asynctimeline.pdf}}%
    \put(0.51481199,0.07573421){\color[rgb]{0.30196078,0.30196078,0.30196078}\makebox(0,0)[lb]{\smash{load}}}%
    \put(0,0){\includegraphics[width=\unitlength,page=4]{asynctimeline.pdf}}%
    \put(0.63360458,0.07573421){\color[rgb]{0.30196078,0.30196078,0.30196078}\makebox(0,0)[lb]{\smash{load}}}%
    \put(0.32120588,0.36463157){\color[rgb]{0,0,0}\makebox(0,0)[lb]{\smash{ }}}%
    \put(0.02545802,0.39994108){\color[rgb]{0,0,0}\makebox(0,0)[lb]{\smash{keep all in memory}}}%
    \put(0,0){\includegraphics[width=\unitlength,page=5]{asynctimeline.pdf}}%
    \put(0.4302456,0.45379373){\color[rgb]{0,0,0}\makebox(0,0)[lb]{\smash{runs out of memory here}}}%
    \put(0.26143674,0.28471184){\color[rgb]{0,0,0}\makebox(0,0)[rb]{\smash{optimal checkpointing}}}%
    \put(0.26022001,0.18453024){\color[rgb]{0,0,0}\makebox(0,0)[rb]{\smash{with asynchronous}}}%
    \put(0,0){\includegraphics[width=\unitlength,page=6]{asynctimeline.pdf}}%
    \put(0.50533438,0.00033422){\color[rgb]{0,0,0}\makebox(0,0)[rb]{\smash{runtime}}}%
    \put(0,0){\includegraphics[width=\unitlength,page=7]{asynctimeline.pdf}}%
    \put(0.26143674,0.25947525){\color[rgb]{0,0,0}\makebox(0,0)[rb]{\smash{(Revolve)}}}%
    \put(0.26022001,0.15929364){\color[rgb]{0,0,0}\makebox(0,0)[rb]{\smash{disk transfer}}}%
    \put(0,0){\includegraphics[width=\unitlength,page=8]{asynctimeline.pdf}}%
    \put(0.89291648,0.43486628){\color[rgb]{0,0,0}\makebox(0,0)[lb]{\smash{forward}}}%
    \put(0.89291648,0.40121749){\color[rgb]{0,0,0}\makebox(0,0)[lb]{\smash{reverse}}}%
    \put(0,0){\includegraphics[width=\unitlength,page=9]{asynctimeline.pdf}}%
  \end{picture}%
\endgroup%

%% file: figures/perfmodel.pdf_tex
\begingroup%
  \makeatletter%
  \providecommand\color[2][]{%
    \errmessage{(Inkscape) Color is used for the text in Inkscape, but the package 'color.sty' is not loaded}%
    \renewcommand\color[2][]{}%
  }%
  \providecommand\transparent[1]{%
    \errmessage{(Inkscape) Transparency is used (non-zero) for the text in Inkscape, but the package 'transparent.sty' is not loaded}%
    \renewcommand\transparent[1]{}%
  }%
  \providecommand\rotatebox[2]{#2}%
  \ifx\svgwidth\undefined%
    \setlength{\unitlength}{379.70056915bp}%
    \ifx\svgscale\undefined%
      \relax%
    \else%
      \setlength{\unitlength}{\unitlength * \real{\svgscale}}%
    \fi%
  \else%
    \setlength{\unitlength}{\svgwidth}%
  \fi%
  \global\let\svgwidth\undefined%
  \global\let\svgscale\undefined%
  \makeatother%
  \begin{picture}(1,0.77758638)%
    \put(0,0){\includegraphics[width=\unitlength]{perfmodel.pdf}}%
    \put(0.03608265,0.02662625){\color[rgb]{0,0,0}\makebox(0,0)[lb]{\smash{0}}}%
    \put(-0.00078851,0.18991281){\color[rgb]{0,0,0}\makebox(0,0)[lb]{\smash{1.25}}}%
    \put(0.03608265,0.67713883){\color[rgb]{0,0,0}\makebox(0,0)[lb]{\smash{5}}}%
    \put(0.07295381,0.00028971){\color[rgb]{0,0,0}\makebox(0,0)[lb]{\smash{1}}}%
    \put(0.44429904,0.00028971){\color[rgb]{0,0,0}\makebox(0,0)[lb]{\smash{1000}}}%
    \put(0.81564427,0.00028971){\color[rgb]{0,0,0}\makebox(0,0)[lb]{\smash{1000000}}}%
    \put(0.18883459,0.7587821){\color[rgb]{0,0,0}\makebox(0,0)[lb]{\smash{Revolve}}}%
    \put(0.42586346,0.7587821){\color[rgb]{0,0,0}\makebox(0,0)[lb]{\smash{I=8}}}%
    \put(0.61021925,0.7587821){\color[rgb]{0,0,0}\makebox(0,0)[lb]{\smash{I=64}}}%
    \put(0.80774331,0.7587821){\color[rgb]{0,0,0}\makebox(0,0)[lb]{\smash{I=1024}}}%
    \put(0.05079681,0.27030581){\color[rgb]{0,0,0}\rotatebox{90}{\makebox(0,0)[lb]{\smash{recompute factor $R$}}}}%
    \put(0.63997806,0.04740582){\color[rgb]{0,0,0}\makebox(0,0)[lb]{\smash{number of layers $n$}}}%
  \end{picture}%
\endgroup%